\documentstyle[preprint,aps]{revtex}

\input epsf         % defines \epsfbox and supporting macros
\epsfverbosetrue    % messages will show height and width
\def\postscript#1{\begin{center}\leavevmode
\hbox{\epsfxsize=0.95\columnwidth\epsfbox{#1}}\end{center}}

\begin{document}

%%%%%%%%%%%%%%%%%%%%%
%\twocolumn[\hsize\textwidth\columnwidth\hsize\csname@twocolumnfalse%
%\endcsname

\draft

\title{C-axis Raman spectra of a normal plane-chain bilayer cuprate
and the pseudogap}

\author{W. C. Wu and J. P. Carbotte}
\address{Department of Physics and Astronomy, McMaster University\\
Hamilton, Ontario, Canada L8S 4M1}
 
\date{\today}
%\date{submitted to PRB in November 1996}

\maketitle

\begin{abstract}
We investigate the Raman spectra in the geometry where  
both incident and scattered photon
polarizations are parallel to the $\hat{\bf z}$-direction, for a 
plane-chain bilayer coupled 
via a single-particle tunneling $t_\perp$. The Raman vertex
is derived in the tight-binding limit and in
the absence of Coulomb screening, the Raman intensity can be separated
into intraband ($\propto t_\perp^4$)
and interband ($\propto t_\perp^2$) transitions.
In the small-$t_\perp$ limit, the interband part dominates and a pseudogap
will appear as it does in the conductivity.
Coulomb interactions bring in a two-particle  
coupling and result in the breakdown of intra- and interband separation. 
Nevertheless, when $t_\perp$ 
is small, the Coulomb screening ($\propto t_\perp^4$) 
has little effect on the intensity to which the unscreened
interband transitions contribute most. In general, the total Raman
spectra are strongly dependent on the magnitude of $t_\perp$.
\end{abstract}

%%%%%%%%%%%%%%%%%%%
%]

\vskip 0.1 true in
\narrowtext

\newpage

\section{Introduction}

Recently, $c$-axis properties of the high-$T_c$ superconductors 
have been of increasing interest both in 
normal and superconducting states. Among several theoretical works 
\cite{CG94,Homes93,Homes95,ZCP,GRS,RL,KJ,Leggett92}, 
a coupled plane-chain model has been proposed \cite{AC96-1,WAC96}
to study various $c$-axis properties of YBaCuO.
Due to a coupling (of magnitude $t_\perp$)
between plane and chain along the $c$-axis,
the properties in the $c$-direction are shown to be quite different as compared
to the ones in the $a$-$b$ plane. More importantly,
a finite band gap (splitting) between the two 
renormalized bands (arising from the finite coupling) 
gives a natural explanation of pseudogaps, which are
mainly observed in the underdoped normal-state copper oxides 
\cite{Basov,PBT96,Hsueh96} and are seen
directly as a depression of the conductivity at
low frequency at low temperatures.

In the past few years, Raman scattering as a probe to study 
high $T_c$ materials have been very successful 
\cite{MPT,BHG,Reznik,CAILH,SNHMV,CLHI,CILH,Dev94-3,HLGTW,HLWG,KR,CITKK94,BC95,BC96} and have in particular played a significant role
in the debate about the symmetry of the gap.
Owing to the layer behavior of the high $T_c$ compounds, most experimental 
and theoretical studies have focused on the $a$-$b$ plane properties, 
taking into account different symmetrical geometries.
In contrast in this paper, we want
specially to study the $c$-axis normal-state Raman scattering,
based on the coupled plane-chain bilayer model.
This not only provides another way to test whether this model
is valid in high $T_c$ cuprates, but also
gives a path for extending calculations from a simple
one band system to the more complicated multiband system.
In particular, a better theory should take Coulomb interactions
into account properly for a multiband system.
  
Devereaux {\em et al.} \cite{DVZ96} have recently given a
detailed account of Raman scattering in 
a superconducting bilayer with two identical planes. They use the
effective mass approximation to derive the Raman vertex and 
have mainly studied the $a$-$b$ plane properties.
In their calculations, they have ignored the interband transition 
which is not so important in the $a$-$b$ plane, but turns out to be 
the dominant part in some cases for the $c$-axis. This case
is discussed in the present paper. 

A summary of what we have done is as follows. We 
have derived the Raman vertex in the 
tight-binding limit which is almost exact in a single bilayer limit.
We have clearly exhibited in the $c$-axis Raman spectra,
the role of the intra- and interband transitions 
which are separable in the absence of the Coulomb screening.
The interband transitions contribute most in the small-$t_\perp$ limit.
While a mixing term comes in when the Coulomb screening is included,
the unscreened interband transition still dominates if $t_\perp$ is small.
Appendix~\ref{appendixa} gives the details of how one can separate
the response functions needed in the calculation of the Raman spectra into 
intraband and interband contributions.
In Appendix~\ref{appendixb}, we present the analogous results 
for the $c$-axis Raman spectra in the case of 
a plane-plane bilayer model. In this instance, a simple analytic end 
expression can be obtained which
helps gain physical insight into the $c$-axis Raman process.
The total volume has been set to one ($\Omega=1$) in this paper.

\section{Formalism}

The non-interacting Hamiltonian for the normal state of a coupled
plane-chain bilayer cuprate is written as
 
\begin{equation}
H_0=\sum_{\bf k}C^\dagger_{\bf k}h({\bf k})C_{\bf k},
\label{eq:H0}
\end{equation}
where we have defined the row and column vectors

\begin{eqnarray}
C^\dagger_{\bf k}=\left(c^\dagger_{1{\bf k}},
                    c^\dagger_{2{\bf k}}\right) ~~~~;~~~~~~~
C_{\bf k}=\pmatrix{c_{1{\bf k}}\cr
                    c_{2{\bf k}}\cr},
\label{eq:c}
\end{eqnarray}
and $c^\dagger_{1{\bf k}},c^\dagger_{2{\bf k}}$ create an electron
in the state ${\bf k}$ in layers 1 and 2 respectively.
The energy matrix

\begin{equation}
h({\bf k})= \pmatrix{\xi_1({\bf k})&t(k_z)\cr
                t(k_z)&\xi_2({\bf k})\cr}.
\label{eq:h}
\end{equation}
Here $\xi_1$ and $\xi_2$ are the (uncoupled) energy bands for the plane and
chain respectively and $t$ (assumed to be real) is the perpendicular 
hopping coupling between plane and chain.  These quantities are defined
explicitly later in Eq.~(\ref{eq:simpleband}).
The spin indices are suppressed throughout this paper.

To study the Raman spectra for a multiband system,
we use an approach which is suitable in the tight-binding limit.
The Hamiltonian (\ref{eq:H0}) can be written in the Wannier 
representation as \cite{AC95}

\begin{eqnarray}
H_0&=&-\sum_{i=1}^2\sum_{{\bf R}_i,{\bf r}_i}
\sigma_i c^\dagger_i({\bf R}_i+{\bf r}_i)c_i({\bf R}_i)\nonumber\\
&-&{t_\perp\over 2}\sum_{{\bf R}_1,{\bf R}_2}
\left\{c_1^\dagger({\bf R}_1)c_2({\bf R}_2)[
\delta_{{\bf R}_1+\hat{\bf z}d/2,{\bf R}_2}+
\delta_{{\bf R}_1-\hat{\bf z}d/2,{\bf R}_2}]+{\rm H.c.} \right\},
\label{eq:H0.tb}
\end{eqnarray}
where ${\bf R}_i$ are the lattice vectors in a plane,
${\bf r}_i$ are the displacements to the nearest neighbors of
${\bf R}_i$, $d$ is the lattice constant
in the $z$ direction, and $\sigma_i$ and $t_\perp$ are the nearest-neighbor 
hopping energies within and between the chain and plane layers. 
The Wannier representation is connected to the
${\bf k}$-space representation by

\begin{eqnarray}
c_i({\bf R}_i)={1\over \sqrt{N}}\sum_{\bf k}e^{i{\bf k}\cdot{\bf R}_i}
c_i({\bf k}),
\label{eq:c.Wannier}
\end{eqnarray}
where $N$ is the total number of lattice sites.
In the presence of a magnetic vector potential
${\bf A}({\bf r})$, the tight-binding Wannier states are modified
by a phase such that \cite{Mahan}

\begin{eqnarray}
c_i({\bf R}_i)\rightarrow c_i({\bf R}_i)
\exp\left[-{ie\over \hbar c}{\bf R}_i \cdot {\bf A}({\bf R}_i)\right],
\label{eq:c.Wannier.phase}
\end{eqnarray}
where $e$ is the charge on the electron,
$\hbar$ is Planck's constant over $2\pi$, and $c$ is the speed of light.
The assumption is made that the vector potential is slowly varying over
the length scale of the crystal lattice and hence
${\bf A}({\bf q})$ is strongly peaked about ${\bf q}=0$.

Substituting (\ref{eq:c.Wannier}) and (\ref{eq:c.Wannier.phase}) 
into (\ref{eq:H0.tb}), to second order in ${\bf A}$, 
Eq.~(\ref{eq:H0.tb}) becomes

\begin{eqnarray}
H=H_0-{e\over 2mc}{\bf p}\cdot {\bf A}+
{e^2\over 2mc^2}{\bf A}\cdot \tensor{\rho}\cdot {\bf A}
\label{eq:H0.A^2}
\end{eqnarray}
in the ${\bf k}$-space, where the vector

\begin{equation}
{\bf p}={m\over \hbar}
\sum_{\bf k}C^\dagger_{\bf k}{\partial h({\bf k})\over 
\partial {\bf k}} C_{\bf k}
\label{eq:p}
\end{equation}
and the tensor

\begin{equation}
\tensor{\rho}={m\over \hbar^2}
\sum_{\bf k}C^\dagger_{\bf k}{\partial^2 h({\bf k})\over 
\partial {\bf k}\partial {\bf k}} C_{\bf k}.
\label{eq:rho.tensor}
\end{equation}
In these formulas $m$ is the bare electron mass.
In electronic Raman scattering, we are interested in
the transition matrix to order ${\bf A}^2$.
Therefore, the term ${\bf A}\cdot\tensor{\rho}\cdot{\bf A}$ in
(\ref{eq:H0.A^2}) is used in first-order perturbation
theory and the term ${\bf p}\cdot {\bf A}$ in (\ref{eq:H0.A^2})
is used in second-order perturbation theory.
As a result, Raman scattering probes an ``effective density'' 

\begin{equation}
\tilde{\rho}\equiv \sum_{\bf k}C^\dagger_{\bf k}\gamma ({\bf k}) C_{\bf k},
\label{eq:rho.tilde}
\end{equation}
where the vertex matrix has the element

\begin{eqnarray}
\gamma_{ij}({\bf k})&=&(\hat{{\bf e}}^I 
\cdot\tensor{\rho}_{ij}\cdot \hat{{\bf e}}^S)
+\sum_{l}\Biggl[
{\langle i,{\bf k}|{\bf p}\cdot \hat{\bf e}^S|l,{\bf k}\rangle
\langle l,{\bf k}|{\bf p}\cdot \hat{\bf e}^I|j,{\bf k}\rangle \over
\epsilon_j({\bf k})-\epsilon_l({\bf k})+\hbar\omega^{I}}\nonumber\\
&+& {\langle i,{\bf k}|{\bf p}\cdot \hat{\bf e}^I|l,{\bf k}\rangle
\langle l,{\bf k}|{\bf p}\cdot \hat{\bf e}^S|j,{\bf k}\rangle \over
\epsilon_j({\bf k})-\epsilon_l({\bf k})-\hbar\omega^{S}}\Biggr].
\label{eq:gamma.original}
\end{eqnarray}
In Eq.~(\ref{eq:gamma.original}),
the summation is over all the uncoupled bands
and the dependence of the momentum transfer ${\bf q}$
on $\gamma_{\bf k}$ is dropped since we assume $q\ll k_F$
(the Fermi momentum).
$\epsilon_i({\bf k})$ is the energy of the $i$-band,
$\hat{\bf e}^I$ ($\hat{\bf e}^S$) is the polarization of
incident (scattered) photon, and the frequency 
$\omega^I$ ($\omega^S$) denotes the incident (scattered) photon energy.

For the problem of a plane-chain bilayer, one has two
bands. In the case that there is a band gap ($\Delta$)
between these two bands which is very small compared to the
energies of incident and scattered photons,
$\Delta\ll \omega^I,\omega^S$, the second term of (\ref{eq:gamma.original})
is negligible. The Raman vertex matrix in (\ref{eq:rho.tilde}) thus
reduces to

\begin{eqnarray}
\gamma({\bf k})={m\over\hbar^2}\sum_{\mu,\nu}\hat{\bf e}^I_\mu
{\partial^2 h({\bf k})\over \partial k_\mu\partial k_\nu}
\hat{\bf e}^S_\nu.
\label{eq:vertex.decomp}
\end{eqnarray}
Eq.~(\ref{eq:vertex.decomp}) shows how one can study various geometries 
for the Raman spectra by choosing the appropriate incident and scattered 
photon polarizations.
We note that while the vertices exhibited in (\ref{eq:vertex.decomp}) 
are somewhat similar to the vertex given by the famous
``effective mass approximation'' \cite{AG74,DVZ96}, the physical
origin is different.

\begin{figure}[h]
\postscript{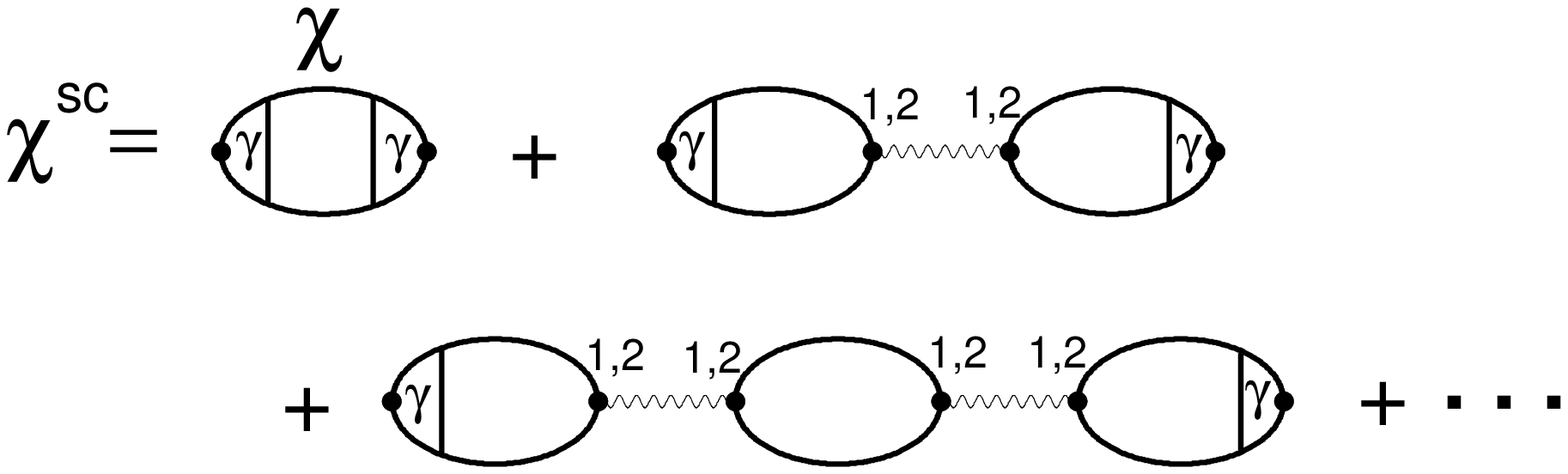}
\vspace{-1cm}
\caption{Diagrams for the screened Raman response function.
Each interaction line comprises four parts corresponding to intralayer
and interlayer Coulomb interactions.}
\label{fig1}
\end{figure}

The differential cross section for Raman scattering is found to be \cite{KD}

\begin{equation}
{d^2\sigma \over d\omega d\Omega} ={1\over \pi}r_0^2 {\omega^S
\over \omega^I}[1+n_B(\omega)]{\rm Im} 
\chi_{\tilde{\rho}\tilde{\rho}}({\bf q}\rightarrow 0,
i\omega_n\rightarrow\omega +i0^+),
\label{eq:Raman.anisotropic}
\end{equation}
where $n_B(\omega)=[\exp(\beta\omega)-1]^{-1}$
is the Bose distribution
function; $r_0=e^2/mc^2$ is the so-called Thompson radius, and
the Raman density response function 
 
\begin{equation}
\chi_{\tilde{\rho}\tilde{\rho}}({\bf q},i\omega_n)=-
\int_0^{\beta}d\tau e^{i\omega_n \tau}
\langle{\rm T}_\tau\tilde{\rho}_{\bf q}(\tau)
\tilde{\rho}_{\bf q}^\dagger(0)\rangle,
\label{eq:chi.rhotilde}
\end{equation}
which is given in terms of the effective density operator $\tilde{\rho}$
given in Eq.~(\ref{eq:rho.tilde}).
Taking into account the screening effect due to the Coulomb energy

\begin{equation}
H_c={1\over 2}\sum_{\bf q}\sum_{i,j=1}^2 U_{ij}({\bf q})
\rho_{i,{\bf q}} \rho_{j,{\bf q}},
\label{eq:Coulomb}
\end{equation}
the Raman response
function (\ref{eq:chi.rhotilde}) in (\ref{eq:Raman.anisotropic})
will be replaced by \cite{KD,MZ,Dev94-1}

\begin{equation}
\chi_{\tilde{\rho}\tilde{\rho}}^{\rm sc}=
\chi_{\tilde{\rho}\tilde{\rho}}
+\sum_{ij=1}^2\chi_{\tilde{\rho}{\rho}_i}U_{ij}({\bf q})
\chi_{{\rho}_j\tilde{\rho}} 
+\sum_{ijml=1}^2\chi_{\tilde{\rho}{\rho}_i}U_{ij}({\bf q})
\chi_{{\rho}_j{\rho}_m}U_{ml}({\bf q})
\chi_{{\rho}_l\tilde{\rho}} +~~\cdot\cdot\cdot
\label{eq:chi.screened}
\end{equation}
Both the intralayer and interlayer Coulomb interactions should be included.
Here the real density on layer $i$ is
 
\begin{equation}
\rho_{i,{\bf q}}=\sum_{\bf k}c^\dagger_{i,{\bf k+q}}c_{i,{\bf k}}
\label{eq:rho_i}
\end{equation}
and the response functions $\chi_{\tilde{\rho}{\rho}_i},
\chi_{{\rho}_i{\rho}_j}$
are defined analogously to $\chi_{\tilde{\rho}\tilde{\rho}}$
by Eq.~(\ref{eq:chi.rhotilde}).
Eq.~(\ref{eq:chi.screened}) corresponds to RPA which is given
diagrammatically in Fig.~\ref{fig1}.

One can easily reduce the infinite series in (\ref{eq:chi.screened}) to

\begin{equation}
\chi_{\tilde{\rho}\tilde{\rho}}^{\rm sc}=
\chi_{\tilde{\rho}\tilde{\rho}}
+(\chi_{\tilde{\rho}{\rho}_1},\chi_{\tilde{\rho}{\rho}_2})(1-U\chi)^{-1}U
\pmatrix{\chi_{{\rho}_1\tilde{\rho}}\cr\chi_{{\rho}_2\tilde{\rho}}\cr},
\label{eq:chi.screened.1}
\end{equation}
where $1$ denotes the $2\times 2$ unit matrix and we have defined
the Coulomb interaction matrix

\begin{eqnarray}
U=\pmatrix{U_{11}&U_{12}\cr U_{21}&U_{22}\cr}
\label{eq:U.Coulomb}
\end{eqnarray}
and the real density response function matrix

\begin{eqnarray}
\chi=\pmatrix{\chi_{\rho_1\rho_1}&\chi_{\rho_1\rho_2}\cr 
\chi_{\rho_2\rho_1}&\chi_{\rho_2\rho_2}\cr}.
\label{eq:chi,matrix}
\end{eqnarray}
In the continuum limit, the intralayer Coulomb interactions
$U_{11}=U_{22}={2\pi e^2\over q_\parallel}$ 
and the interlayer Coulomb interactions
$U_{12}=U_{21}={2\pi e^2\over q_\parallel}e^{-q_\parallel c}$ 
($q_\parallel$ is the momentum transfer parallel to $a$-$b$ plane and
$c$ is the bilayer spacing \cite{GP}) which 
all go to infinity in the strong screening ${\bf q}_\parallel
\rightarrow 0$ limit. In the tight-binding strong-coupling limit, the 
couplings $U_{ij}$ are themselves large. In either cases, one 
can approximate (\ref{eq:chi.screened.1}) by

\begin{equation}
\chi_{\tilde{\rho}\tilde{\rho}}^{\rm sc}=
\chi_{\tilde{\rho}\tilde{\rho}}
-(\chi_{\tilde{\rho}{\rho}_1},\chi_{\tilde{\rho}{\rho}_2})\chi^{-1}
\pmatrix{\chi_{{\rho}_1\tilde{\rho}}\cr\chi_{{\rho}_2\tilde{\rho}}\cr}.
\label{eq:chi.screened.2}
\end{equation}
The minus sign in (\ref{eq:chi.screened.2}) clearly exhibits
the screening effect of the Coulomb interaction.

\section{Results and Discussions}
 
Eqs.~(\ref{eq:Raman.anisotropic}) and (\ref{eq:chi.screened.2}) allow
one to study various symmetries of the {\em screened} Raman intensities
for two-band systems. In this paper, however, 
we will focus on the $c$ (or $z$) axis Raman
intensities. We first specify the two uncoupled bands and the coupling:
 
\begin{eqnarray}
\xi_1(k_x,k_y)&=&{\hbar^2\over 2m}(k_x^2 +k_y^2)-\mu+\Delta\nonumber\\
\xi_2(k_y)&=&{\hbar^2 \over 2m}k_y^2-\mu\nonumber\\
t(k_z)&=&-2t_\perp \cos(k_z d/2),
\label{eq:simpleband}
\end{eqnarray}
where $\Delta$ corresponds to the band splitting (responsible
for the ``pseudogap''), $\mu$ is the chemical potential, and
$t_\perp$ is the plane-chain coupling strength.
While the bands $\xi_1$ and $\xi_2$ with circular and linear Fermi surface
are used to simplify the calculation, they give
qualitatively similar results to those given by more realistic
tight-binding bands \cite{AC96-1}.
Since the $k_z$ dependence only comes in through $t(k_z)$ 
which is assumed to have no $k_x$ and $k_y$ dependence, the only relevant
geometry for the $c$-axis Raman spectra will be
$(\hat{\bf e}^S,\hat{\bf e}^I)=(\hat{\bf z},\hat{\bf z})$.
Following (\ref{eq:vertex.decomp}), we define
 
\begin{eqnarray}
\gamma_c(k_z)\equiv\pmatrix{0&\gamma_{zz}\cr\gamma_{zz}&0\cr},
\label{eq:gamma_z}
\end{eqnarray}
where $\gamma_{zz}=(m/\hbar^2)\partial^2 t(k_z)/\partial k_z^2
=(t_\perp d^2m/2\hbar^2)\cos(k_z d/2)$.
Eq.~(\ref{eq:gamma_z}) is the Raman vertex used in (\ref{eq:rho.tilde}) for
calculating the ``$c$-axis Raman intensities'' throughout this paper.
 
In Appendix~\ref{appendixa}, we have shown how one can separate
all the response functions of Eq.~(\ref{eq:chi.screened.2})
(needed in the calculation of the $c$-axis Raman spectra)
into an {\em intraband} and an {\em interband} contributions.
With these separations, one can easily estimate how each
susceptibility $\chi$ depends on the coupling strength $t_\perp$, 
for both intraband and interband transitions.
Of most interest, one sees that the unscreened Raman
response function $\chi_{\tilde{\rho}\tilde{\rho}}$ has an intraband
contribution which is proportional $t_\perp^4$ and an interband
contribution which is proportional $t_\perp^2$. Therefore,
the interband transitions dominate for the unscreened 
$c$-axis Raman intensities when $t_\perp$ is small.

In Fig.~\ref{fig2}, we first show the {\em unscreened} Raman intensities for 
small $t_\perp=2{\rm meV}$ at different temperatures.
Fixed parameters $\Delta=20{\rm meV}$ and $\mu=500{\rm meV}$ are used
for all the results presented in this paper. 
To include the impurity scattering, we then introduce the scattering rates
$\Gamma_i$ into the Green's functions in (\ref{eq:G0}) 

\begin{eqnarray}
G_0^{-1}({\bf k},i\omega_n)=\pmatrix{
i\omega_n-\xi_1+i\Gamma_1{\rm sgn}(\omega_n) &-t\cr 
-t&i\omega_n-\xi_2+i\Gamma_2{\rm sgn}(\omega_n)\cr}.
\label{eq:G0.damped}
\end{eqnarray}
For simplicity, we will assume that the impurity scattering rates
are equal for both bands ($\Gamma_1=\Gamma_2\equiv\Gamma$)
and $\Gamma$ is linear in temperature ($\Gamma=20{\rm meV}$
at $T=100K$), but independent of frequency and momentum.
As seen in Fig.~\ref{fig2}, the $c$-axis Raman spectra can exhibit
clear signatures of a pseudogap, as long as
the temperature is not too high. At high temperatures for which
$\Gamma\geq \Delta$, the pseudogap feature is washed out
by the impurity broadening. This is, in particular, shown by
the different low-frequency dependence of the intensities
(see the inset in Fig.~\ref{fig2}). 

In Fig.~\ref{fig3}, we present the $T=20K$ unscreened Raman intensities for
different coupling strengths. The inset shows that the unscreened 
$c$-axis Raman intensity can, in general, be separated into
an intraband and an interband contribution.
For better clarity of presentation, all intensities 
have been divided by $t_\perp^2$. 
This leads to a almost constant interband contribution 
plus a intraband contribution proportional to
$t_\perp^2$. One finds at larger
$t_\perp$ that the pseudogap feature arising from the interband transitions
are overwhelmed by the increasing contribution of
intraband transitions which tend to dominate the Raman response. 

We next consider the {\em screened} Raman intensity based
on the full prescription of Eq.~(\ref{eq:chi.screened.2}).
Due to the Coulomb screening, the presence of the second (mixing) 
term in (\ref{eq:chi.screened.2})
results in the breakdown of the intraband and interband separation
for the unscreened Raman intensity. For comparison with Figs.~\ref{fig2} and 
\ref{fig3}, we have presented in Figs.~\ref{fig4} and \ref{fig5}
the screened $c$-axis Raman intensities.
Comparing Fig.~\ref{fig4} with Fig.~\ref{fig2}, one sees no major 
difference between the unscreened and screened intensities 
except for different spectral weights of the higher-frequency intensity. 
This result is valid for the small-$t_\perp$ case and can be established
as follows. When $t_\perp$ is small, one can ignore $\chi_{{\rho}_1{\rho}_2}$ 
$\sim t_\perp^2$ in $\chi$ in (\ref{eq:chi,matrix}) as
compared to $\chi_{{\rho}_1{\rho}_1},\chi_{{\rho}_2{\rho}_2}$
$\sim t_\perp^0$ (see Appendix~\ref{appendixa}).  Consequently, 
(\ref{eq:chi.screened.2}) can be reduced to a more elegant result
 
\begin{equation}
\chi_{\tilde{\rho}\tilde{\rho}}^{\rm sc}=
\chi_{\tilde{\rho}\tilde{\rho}}
-\sum_{i=1}^2 {(\chi_{\tilde{\rho}{\rho}_i})^2\over
\chi_{{\rho}_i {\rho}_i}}.
\label{eq:chi.screened.3}
\end{equation}
That is, the screening effects can be separated for each layer.
However, the complete separation of
intraband and interband contributions is still not possible
because the intra- and interband parts (both $\sim t_\perp^2$) of
$\chi_{\tilde{\rho}{\rho}_i}$ are equally important
(although $\chi_{{\rho}_i {\rho}_i}\sim t_\perp^0$ is dominated by the
intraband transition). Nevertheless, the mixing (second) term in 
(\ref{eq:chi.screened.3}) is proportional to $t_\perp^4$ and has only a 
small effect when $t_\perp$ is small.  Therefore, the unscreened 
interband transition given by $\chi_{\tilde{\rho}\tilde{\rho}}$ 
still dominates.
For higher values of $t_\perp$, as used in Figs.~\ref{fig5} and
\ref{fig3}, the Coulomb screening has a large effect. In fact,
in view of the inset in Fig.~\ref{fig5}, 
the  Coulomb screening effect tends to cancel against the unscreened
interband contribution, largely leaving the unscreened intraband transition.

\section{Conclusions}

In this paper, we have studied the $c$-axis Raman spectra such that
both the incident and scattered photon polarizations are parallel to 
the $\hat{\bf z}$-direction, for a plane-chain bilayer two-band cuprate. 
This coupled plane-chain model \cite{AC96-1} has given a natural 
explanation of pseudogaps \cite{Basov,PBT96,Hsueh96} seen in the AC $c$-axis 
optical conductivity. Here we have shown in detail how one  
can also study pseudogaps by doing $c$-axis Raman scattering. 
Our calculations are based on a tight-binding model for the Raman vertex. 
Coulomb screening effects are treated properly and
include both intralayer and interlayer Coulomb interactions.  

When the Coulomb screening is absent, the Raman intensity can be
separated into an intraband and an interband contribution.
The interband contribution ($\propto t_\perp^2$) dominates
in the small-$t_\perp$ limit. However, the presence of the Coulomb
screening mixes up this separation and the resulting intensities are 
strongly dependent on the magnitude of $t_\perp$.
Nevertheless, when $t_\perp$ is small, the mixing effect arising from 
Coulomb screening is small and, as a result,
the interband unscreened transition still dominates.

It's obvious that the approach derived in this
paper for calculating Raman spectra in a two-band system can be 
easily extended to the case of a multiband system.
A extension of our calculations of $c$-axis Raman intensities to
the case of a superconducting plane-chain bilayer cuprate will be
given elsewhere including the effect of impurities.
 
\acknowledgments
We thank W.A. Atkinson and D. Branch for useful discussions.
This work was supported by Natural Sciences and Engineering Research Council
(NSERC) of Canada and Canadian Institute for Advanced Research (CIAR).

\appendix
\section{Intraband and Interband Transitions}
\label{appendixa}

In this appendix, we show how one can 
separate the response functions needed in the 
calculation of the Raman spectra into intraband and interband
transitions. Consider a  general response function 

\begin{equation}
\chi_{AB}({\bf q},i\omega_n)=-
\int_0^{\beta}d\tau e^{i\omega_n \tau}
\langle{\rm T}_\tau A_{\bf q}(\tau)
B_{\bf q}^\dagger(0)\rangle,
\label{eq:chi.app}
\end{equation}
where here, the operators $A,B$ can be either the effective 
density $\tilde{\rho}$ or the real densities $\rho_1$ or $\rho_2$.
These operators can be written in a unified form such that in the limit 
${\bf q}\rightarrow 0$

\begin{equation}
A= \sum_{\bf k} C^\dagger_{{\bf k}} \gamma({\bf k})
C_{{\bf k}}=\sum_{\bf k}\sum_{i,j=1}^2\gamma^{A}_{ij}c^\dagger_{i,{\bf k}}
c_{j,{\bf k}}.
\label{eq:rho.app}
\end{equation}
In (\ref{eq:rho.app}), the matrix $\gamma$ is $\gamma_c$
given in (\ref{eq:gamma_z}) for $\tilde{\rho}$ and  

\begin{eqnarray}
\gamma_1=\pmatrix{1&0\cr0&0\cr}~~\mbox{for ~$\rho_1$}~~;~~~~~ 
\gamma_2=\pmatrix{0&0\cr0&1\cr}~~\mbox{for ~$\rho_2$}.
\label{eq:gamma12}
\end{eqnarray}
Using (\ref{eq:rho.app}), the average $\langle\cdot\cdot\cdot\rangle$
in (\ref{eq:chi.app}) becomes

\begin{eqnarray}
&&\langle{\rm T}_\tau A_{\bf q}(\tau) B_{\bf q}^\dagger(0)\rangle
\nonumber\\
&&=\sum_{ijml=1}^2\sum_{\bf k}\sum_{\bf k^\prime}
\gamma_{ij}^A ({\bf k}) \gamma_{ml}^B ({\bf k^\prime})
\langle{\rm T}_\tau c_{i,{\bf k}}^\dagger(\tau)
c_{j,{\bf k}}(\tau) c_{m,{\bf k^\prime}}^\dagger(0)
c_{l,{\bf k^\prime}}(0)\rangle\nonumber\\
&&=-\sum_{ijml=1}^2\sum_{\bf k}\sum_{\bf k^\prime}
\gamma_{ij}^A ({\bf k}) \gamma_{ml}^B ({\bf k^\prime})
\langle{\rm T}_\tau c_{l,{\bf k}}(0)c^\dagger_{i,{\bf k^\prime}}(\tau)\rangle
\langle{\rm T}_\tau c_{j,{\bf k}}(\tau)c^\dagger_{m,{\bf k^\prime}}(0)\rangle
\nonumber\\
&&=-\sum_{ijml=1}^2\sum_{\bf k}\gamma_{ij}^A ({\bf k})
\gamma_{ml}^B ({\bf k})G^0_{li}({\bf k},-\tau)G^0_{jm}({\bf k},\tau).
\label{eq:expand.app}
\end{eqnarray}
The development from the second to the third line in 
(\ref{eq:expand.app}) makes use of the usual 
Hartree-Fock approximation. $G^0_{ij}$ are the elements of a 
$2\times 2$ non-interacting Green's function matrix defined by

\begin{equation}
G_0({\bf k},\tau)= 
-\langle{\rm T}_\tau C_{\bf k}(\tau)C^\dagger_{\bf k}(0) \rangle.
\label{eq:G0.def}
\end{equation}
For the present coupled plane-chain bilayer model,
one has (see (\ref{eq:h}))

\begin{eqnarray}
G_0^{-1}({\bf k},i\omega_n)=\pmatrix{i\omega_n-\xi_1&-t\cr
                        -t&i\omega_n-\xi_2\cr}.
\label{eq:G0}
\end{eqnarray}
Substituting (\ref{eq:expand.app}) into (\ref{eq:chi.app}) and making
use of the Fourier transform

\begin{equation}
G^0_{ij}({\bf k},\tau)={1\over \beta}\sum_{i\omega_{n^\prime}}
e^{-i\omega_{n^\prime}\tau}G^0_{ij}({\bf k},i\omega_{n^\prime})
\label{eq:G.F.T.}
\end{equation}
and the orthogonal relation

\begin{equation}
{1\over \beta}\int_{0}^{\beta}d\tau~e^{i(\omega_n -\omega_{n^\prime})\tau}=
\delta_{\omega_n,\omega_{n^\prime}},
\label{eq:ortho}
\end{equation}
one obtains

\begin{equation}
\chi_{AB}({\bf q}\rightarrow 0,i\omega_n)=
{1\over \beta}\sum_{\omega_{n^\prime}}\sum_{\bf k}
\sum_{ijml=1}^2 \gamma^A_{ij}({\bf k}) \gamma^B_{ml}({\bf k})
G^0_{li}({\bf k},i\omega_{n^\prime})G^0_{jm}({\bf k},
i\omega_{n^\prime}+i\omega_n).
\label{eq:chi.app.1}
\end{equation}
Furthermore, using (\ref{eq:gamma_z}), (\ref{eq:gamma12}), and 
the fact that $G^0_{12}=G^0_{21}$, one finds

\begin{eqnarray}
\chi_{\tilde{\rho}\tilde{\rho}}&=&
{1\over \beta}\sum_{\omega_{n^\prime}}\sum_{\bf k}
{\rm Tr} [G_0({\bf k},i\omega_{n^\prime})\gamma_c
G_0({\bf k},i\omega_{n^\prime}+i\omega_n)\gamma_c]\nonumber\\
\chi_{\tilde{\rho}{\rho}_i}&=&\chi_{{\rho}_i\tilde{\rho}}=
{1\over \beta}\sum_{\omega_{n^\prime}}\sum_{\bf k}
{\rm Tr} [G_0({\bf k},i\omega_{n^\prime})\gamma_c
G_0({\bf k},i\omega_{n^\prime}+i\omega_n)\gamma_i]\nonumber\\
\chi_{{\rho}_i{\rho}_j}&=&\chi_{{\rho}_j{\rho}_i}=
{1\over \beta}\sum_{\omega_{n^\prime}}\sum_{\bf k}
{\rm Tr}[G_0({\bf k},i\omega_{n^\prime})\gamma_i
G_0({\bf k},i\omega_{n^\prime}+i\omega_n)\gamma_j].
\label{eq:various.res.fcns}
\end{eqnarray}
The trace operator enables one to calculate 
$\chi_{\tilde{\rho}\tilde{\rho}}$, etc in (\ref{eq:various.res.fcns})
in any rotated frame which is convenient. We will choose a frame in which
the Green's function matrix is diagonal. This leads to results that
are readily interpretable physically (see later).

We first diagonalize the Hamiltonian $h$ in (\ref{eq:h})
and the Green's function matrix
$G_0^{-1}$ in (\ref{eq:G0})
 
\begin{eqnarray}
h({\bf k})\rightarrow \hat{h}({\bf k})=U^\dagger({\bf k}) h({\bf k})U({\bf k})
=\pmatrix{\epsilon_+&0\cr
          0&\epsilon_-\cr},\nonumber\\
G_0^{-1}\rightarrow \hat{G}_0^{-1}=U^\dagger G_0^{-1}U
=\pmatrix{i\omega_n-\epsilon_+&0\cr
0&i\omega_n-\epsilon_-\cr}, \raisebox{1.0cm}{~~~}
\label{eq:diag.G}
\end{eqnarray}
where the unitary matrix is
 
\begin{eqnarray}
U({\bf k})= {1\over \sqrt{\epsilon_+ - \epsilon_-}}\pmatrix{
        -{t\over |t|}\sqrt{\xi_1-\epsilon_-}&
        -{t\over |t|}\sqrt{\epsilon_+ -\xi_1}\cr
        -\sqrt{\epsilon_+ -\xi_1}&\sqrt{\xi_1-\epsilon_-}\cr}
\label{eq:U}
\end{eqnarray}
and the eigenvalues are
 
\begin{equation}
\epsilon_\pm={\xi_1+\xi_2\over 2}\pm
\sqrt{\left({\xi_1-\xi_2\over 2}\right)^2+t^2}.
\label{eq:spectrum2}
\end{equation}
In an analogous way, one can rotate the vertices
 
\begin{eqnarray}
\gamma_c\rightarrow \hat{\gamma}_c=U^\dagger \gamma_c U
={\gamma_{zz}\over\epsilon_+ -\epsilon_-}
\pmatrix{2t&-{t\over |t|}(\xi_1 -\xi_2)\cr
          -{t\over |t|}(\xi_1 - \xi_2) &-2t\cr},
\label{eq:diag.gamma_c}
\end{eqnarray}

\begin{eqnarray}
\gamma_1\rightarrow \hat{\gamma}_1=U^\dagger \gamma_1 U
={1\over\epsilon_+ -\epsilon_-}
\pmatrix{\xi_1-\epsilon_-&|t|\cr
          |t| &\epsilon_+ -\xi_1\cr},
\label{eq:diag.gamma_1}
\end{eqnarray}
and

\begin{eqnarray}
\gamma_2\rightarrow \hat{\gamma}_2=U^\dagger \gamma_2 U
={1\over\epsilon_+ -\epsilon_-}
\pmatrix{\epsilon_+ -\xi_1&-|t|\cr
          -|t| &\xi_1-\epsilon \cr}.
\label{eq:diag.gamma_2}
\end{eqnarray}
For convenience, we redefine

\begin{eqnarray}
&&\hat{G}_0=
\pmatrix{G_{11}&0\cr 0&G_{22}\cr},~~~~~
\hat{\gamma}_c=
\pmatrix{\gamma_{11}&\gamma_{12}\cr\gamma_{12}&-\gamma_{11}\cr},\nonumber\\
&&\hat{\gamma}_1=
\pmatrix{\alpha_{11}&\alpha_{12}\cr\alpha_{12}&\alpha_{22}\cr},~~~~~
\hat{\gamma}_2=
\pmatrix{\alpha_{22}&-\alpha_{12}\cr -\alpha_{12}&\alpha_{11}\cr},
\raisebox{1.0cm}{~~~}
\label{eq:diag.gamma_all.new}
\end{eqnarray}
where one can easily identify these new variables. 
Substituting (\ref{eq:diag.gamma_all.new}) into 
(\ref{eq:various.res.fcns}), we obtain

\begin{eqnarray}
\chi_{\tilde{\rho}\tilde{\rho}}&=&
{1\over \beta}\sum_{\omega_{n^\prime}}\sum_{\bf k}
\gamma_{11}^2\Bigl(G_{11}G_{11}^\prime+ G_{22}G_{22}^\prime \Bigr)+
\gamma_{12}^2\Bigl(G_{11}G_{22}^\prime+ G_{22}G_{11}^\prime \Bigr),
\nonumber\\
\chi_{\tilde{\rho}{\rho}_1}&=&\chi_{{\rho}_1\tilde{\rho}}=
{1\over \beta}\sum_{\omega_{n^\prime}}\sum_{\bf k}
\gamma_{11}\Bigl(\alpha_{11}G_{11}G_{11}^\prime-\alpha_{22}
G_{22}G_{22}^\prime \Bigr)+
\gamma_{12}\alpha_{12}\Bigl(G_{11}G_{22}^\prime+ G_{22}G_{11}^\prime \Bigr),
\nonumber\\
\chi_{\tilde{\rho}{\rho}_2}&=&\chi_{{\rho}_2\tilde{\rho}}=
{1\over \beta}\sum_{\omega_{n^\prime}}\sum_{\bf k}
\gamma_{11}\Bigl(\alpha_{22}G_{11}G_{11}^\prime-\alpha_{11}
G_{22}G_{22}^\prime \Bigr)-
\gamma_{12}\alpha_{12}\Bigl(G_{11}G_{22}^\prime+ G_{22}G_{11}^\prime \Bigr),
\nonumber\\
\chi_{{\rho}_1{\rho}_1}&=&
{1\over \beta}\sum_{\omega_{n^\prime}}\sum_{\bf k}
\Bigl(\alpha_{11}^2 G_{11}G_{11}^\prime+\alpha_{22}^2 
G_{22}G_{22}^\prime \Bigr)+
\alpha_{12}^2\Bigl(G_{11}G_{22}^\prime+ G_{22}G_{11}^\prime \Bigr),
\nonumber\\
\chi_{{\rho}_2{\rho}_2}&=&
{1\over \beta}\sum_{\omega_{n^\prime}}\sum_{\bf k}
\Bigl(\alpha_{22}^2 G_{11}G_{11}^\prime+\alpha_{11}^2 
G_{22}G_{22}^\prime \Bigr)+
\alpha_{12}^2\Bigl(G_{11}G_{22}^\prime+ G_{22}G_{11}^\prime \Bigr),
\nonumber\\
\chi_{{\rho}_1{\rho}_2}&=&\chi_{{\rho}_2{\rho}_1}=
{1\over \beta}\sum_{\omega_{n^\prime}}\sum_{\bf k}
\alpha_{11}\alpha_{22}\Bigl(G_{11}G_{11}^\prime+G_{22}G_{22}^\prime \Bigr)-
\alpha_{12}^2\Bigl(G_{11}G_{22}^\prime+ G_{22}G_{11}^\prime \Bigr),
\label{eq:various.res.fcns.1}
\end{eqnarray}
where $G\equiv G({\bf k},i\omega_{n^\prime})$ and
$G^\prime\equiv G({\bf k},i\omega_{n^\prime}+i\omega_{n})$.
Clearly for each $\chi$ in (\ref{eq:various.res.fcns.1}),
the first term corresponds
to an {\em intraband} contribution and the second term corresponds
to an {\em interband} contribution.
Crudely speaking,  when $t_\perp$ is small
(recall (\ref{eq:simpleband}) and (\ref{eq:gamma_z})),
$\gamma_{11}\sim t_\perp^2$, $\gamma_{12}\sim t_\perp^1$,
$\alpha_{11}\sim t_\perp^0$, $\alpha_{22}\sim t_\perp^2$,
and $\alpha_{12}\sim t_\perp^1$.
These in turn allow one to estimate the $t_\perp$-dependence of
each $\chi$, for both intraband and interband transitions.

\section{Plane-Plane Bilayer Model}
\label{appendixb}
 
In this Appendix, we present
the analogous results for the $c$-axis Raman
intensities in a coupled plane-plane bilayer. Due to the identity of these
two layers, analytical results are available and hence 
gives more insight into the numerical results of the plane-chain model.  
Similar to Eq.~(\ref{eq:simpleband}), we use the simple bands

\begin{eqnarray}
\xi_1(k_x,k_y)={\hbar^2\over 2m}(k_x^2 +k_y^2)-\mu+\Delta\nonumber\\
\xi_2(k_x,k_y)={\hbar^2 \over 2m}(k_x^2+k_y^2)-\mu,
\label{eq:simpleband.pp}
\end{eqnarray}
for the two planes which are separated by a band gap $\Delta$ (pseudogap).
The coupling between these two planes is the same as given 
in (\ref{eq:simpleband}).
In the case of small $t_\perp$, one can approximate 
$\epsilon_+=\xi_1$ and  $\epsilon_-=\xi_2$ (see (\ref{eq:spectrum2})).

\begin{figure}[h]
\postscript{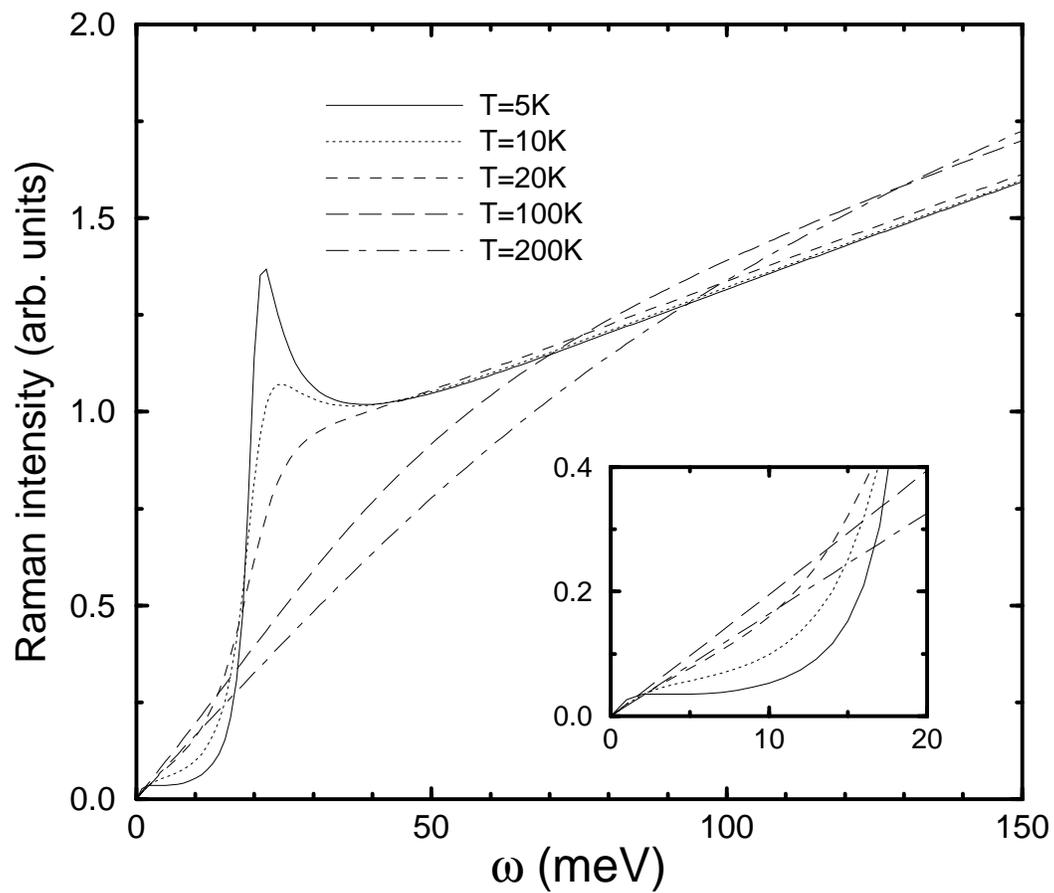}
\vspace{-1cm}
\caption{Unscreened $c$-axis Raman intensities for $t_\perp=2{\rm meV}$
at various temperatures. We magnify in the inset 
the different $\omega$-dependence of the low-frequency intensities.}
\label{fig2}
\end{figure}

\begin{figure}[h]
\postscript{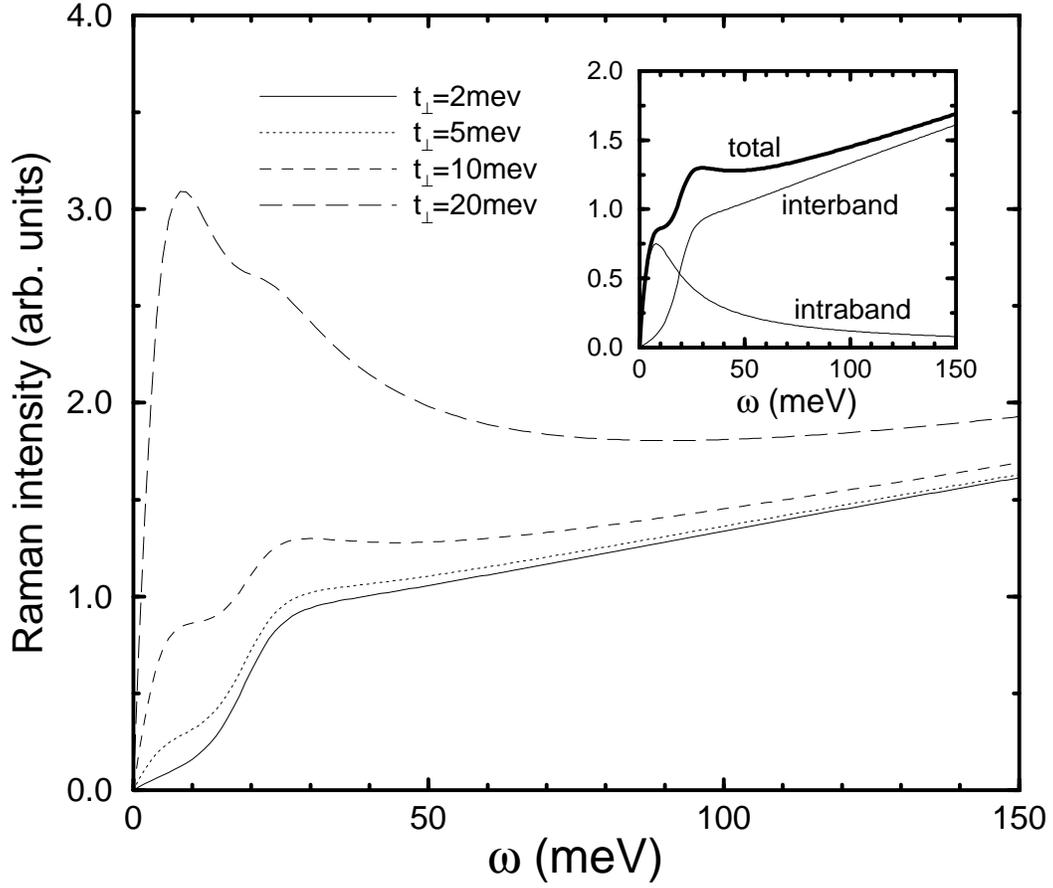}
\caption{Unscreened $c$-axis Raman intensities at $T=20K$ with
different values of plane-chain coupling $t_\perp$.
For easier comparison, all intensities are divided by $t_\perp^2$.
The inset shows the separation into 
intraband and interband transitions of the unscreened $c$-axis 
Raman intensity (for the $t_\perp=10{\rm meV}$ case).}
\label{fig3}
\end{figure}

\begin{figure}[h]
\postscript{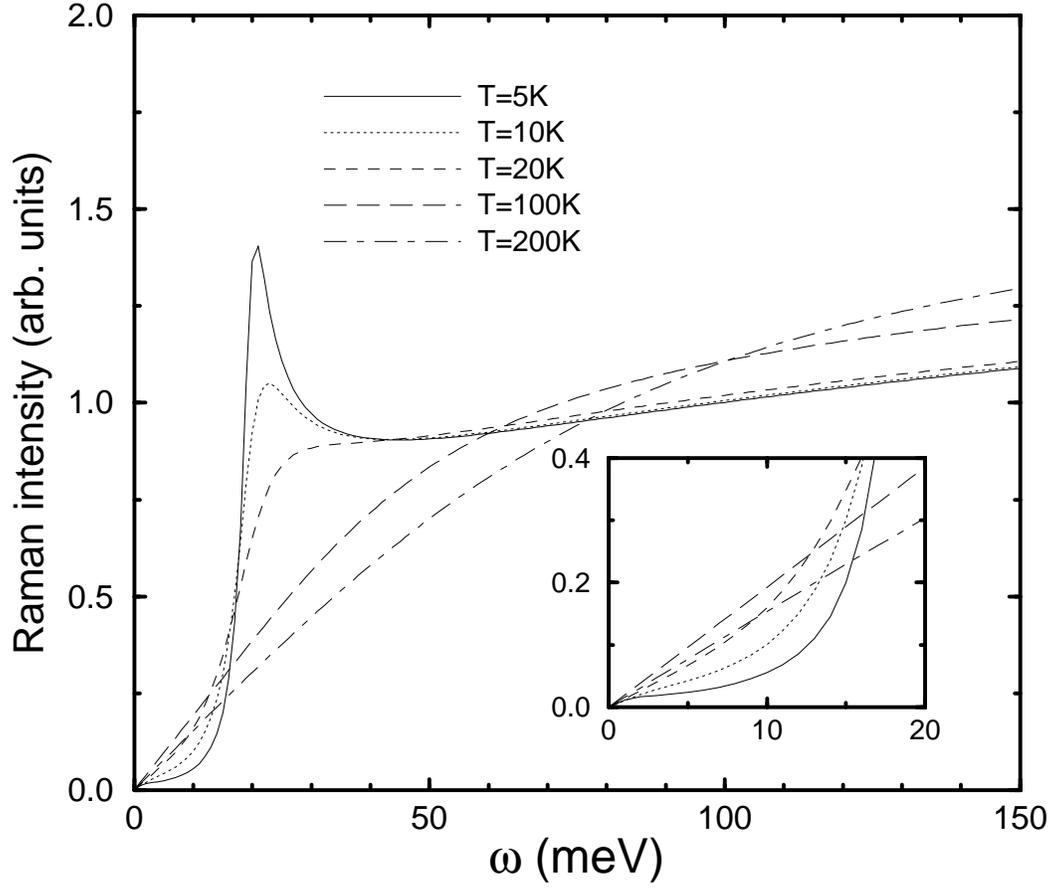}
\vspace{-1cm}
\caption{Screened $c$-axis Raman intensities for $t_\perp=2{\rm meV}$
at various temperatures (to be compared to Fig.~\protect\ref{fig2}).}
\label{fig4}
\end{figure}

\begin{figure}[h]
\vspace{-1cm}
\postscript{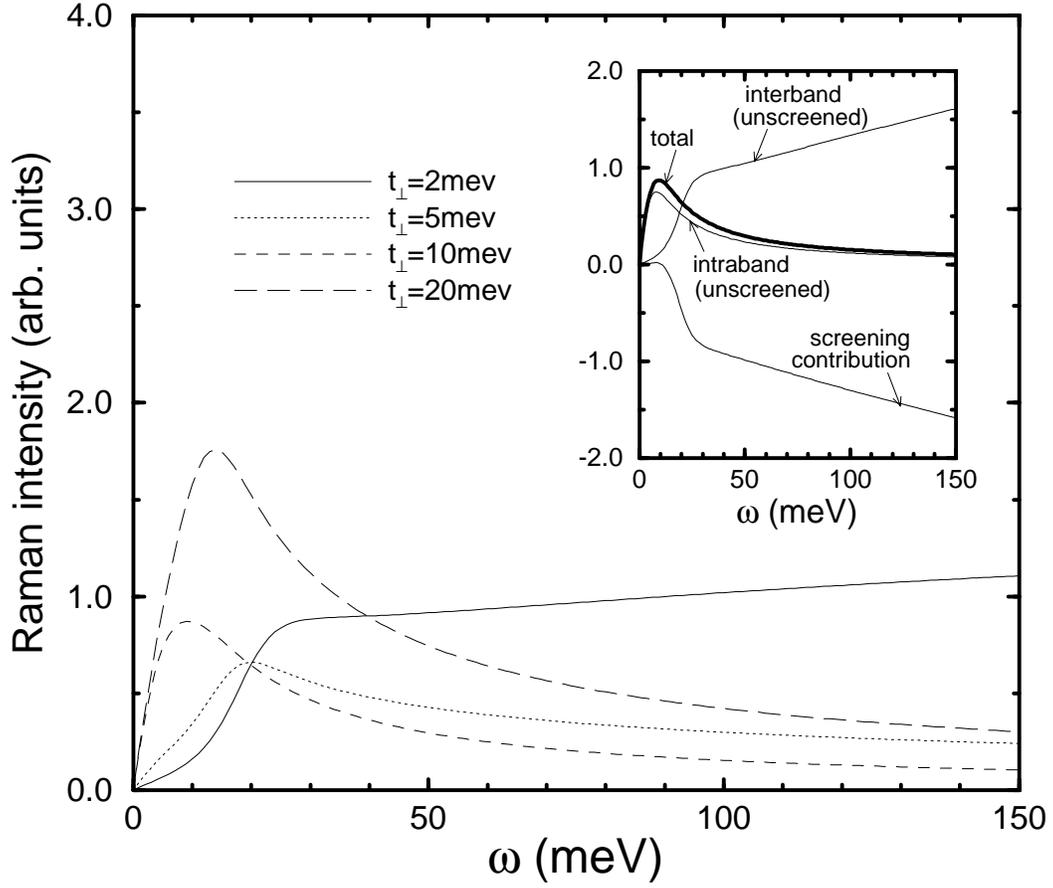}
\vspace{-1cm}
\caption{Screened  $c$-axis Raman intensities at $T=20K$
for various values of $t_\perp$ (to be compared to Fig.~\protect\ref{fig3}).
The inset (for $t_\perp=10{\rm meV}$) shows separably the 
three different contributions, namely the unscreened intraband, 
unscreened interband, and screening effect.}
\label{fig5}
\end{figure}

\begin{figure}[h]
\postscript{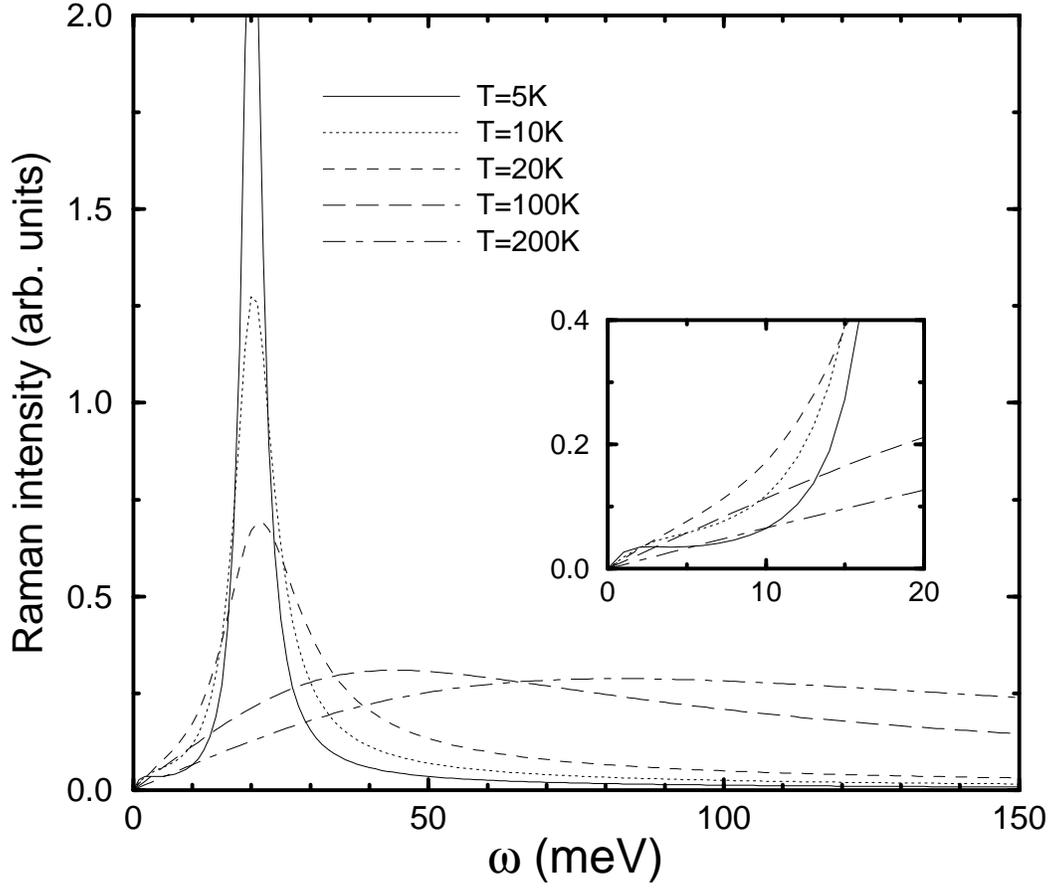}
\vspace{-1cm}
\caption{Unscreened $c$-axis Raman intensities in a coupled plane-plane
bilayer model for $t_\perp=2{\rm meV}$
at various temperatures. The inset
shows the different low-frequency $\omega$-dependence of the intensities.
This is to be compared with Fig.~\protect\ref{fig2} which applies
to the plane-chain bilayer.}  
\label{fig6}
\end{figure}

With the above simplification, the unscreened $c$-axis Raman intensity 
given by the imaginary part of
$\chi_{\tilde{\rho}\tilde{\rho}}$  in (\ref{eq:various.res.fcns.1}) 
can easily be worked out to be proportional to

\begin{eqnarray}
t_\perp^2 \omega \left[{6t_\perp^2\over\Delta^2}
{4\Gamma\over (\hbar\omega)^2+4\Gamma^2}
+\left({2\Gamma\over (\hbar\omega+\Delta)^2+4\Gamma^2}+
{2\Gamma\over (\hbar\omega-\Delta)^2+4\Gamma^2}\right)\right],
\label{eq:raman.total.pp}
\end{eqnarray}
where the first term corresponds to the familiar intraband transition 
($\propto t_\perp^4$) and the
second term corresponds to the interband transition ($\propto t_\perp^2$).
One can see from (\ref{eq:raman.total.pp}) that
apart from the prefactor $t_\perp^2 \omega$,
the interband term contains the same 
factors as the intraband term except for a peak shift
from $\omega=0$ to $\omega=\pm \Delta$ in the Lorentzian form. 
At zero frequency the first Drude like contribution in 
(\ref{eq:raman.total.pp}) will go like $6t_\perp^2/\Delta^2 \Gamma$
while the second (interband) term will go instead like
$4\Gamma/\Delta^2$ which is the opposite dependence
on $\Gamma$ than exhibited by the first term.
Also for $t_\perp\rightarrow 0$ the Drude term is small compared 
with the second term which will then exhibit a pseudogap for 
$\hbar\omega\alt \Delta$ as $\Gamma\rightarrow 0$.
Using (\ref{eq:raman.total.pp}), we plot in Fig.~\ref{fig6}
the temperature-dependent $c$-axis
Raman intensities for the plane-plane bilayer model.
All the parameters used are the same as those used in Fig.~\ref{fig2}.  
It is seen in Fig.~\ref{fig6} that
a pseudogap can develop in the $c$-axis Raman
intensities at low frequencies provided the
temperature (or the impurity damping rate $\Gamma$)
is not too high. We remark that, in analogy to the plane-chain case, 
the Coulomb screening has little effect on the small-$t_\perp$
unscreened intensities given in Fig.~\ref{fig6}.

The major difference between the results in Fig.~\ref{fig2}
for a plane-chain bilayer and those in Fig.~\ref{fig6}
for a plane-plane bilayer is that the spectral weight just beyond the
pseudogap frequency is much higher in the former case than in the latter case.
This can be explained as follows.
For a plane-plane bilayer in which the two bands are separated
by a constant band gap ($\Delta$) all over the brillouin zone, the interband 
transitions will be peaked at $\omega=\Delta$.
In contrast in the plane-chain bilayer,  the band gap between the
plane and chain bands varies at different ${\bf k}$ points (with minimum
value of $\Delta$) and, as a result, the interband transitions 
are spreaded out for $\omega\geq\Delta$.

\end{document}